\documentclass[12pt]{article}
\usepackage{amsmath,amssymb,amsfonts}

\newtheorem{teo}{Theorem}
\newtheorem{ex}{Example}
\newtheorem{de}{Definition}
\newtheorem{lem}{Lemma}
\newtheorem{rem}{Remark}
\newtheorem{pro}{Proposition}

\def\d{\partial}
\def\n{\noindent}
\def\f{\frac}

\begin{document}

\title{Flat bidifferential ideals\\
 and semihamiltonian PDEs
\footnote{
Appeared in Journal of Physics A: Mathematical and General 39 (2006) 
13701--13715,\hspace{1 cm} and  available at 
http://www.iop.org/EJ/abstract/-search=14237525.1/0305-4470/39/44/006,
\hspace{1 cm}
\copyright copyright (2006) IOP Publishing Ltd.
}
}                     
\author{Paolo Lorenzoni\\
Dipartimento di Matematica e Applicazioni\\
Universit\`a di Milano-Bicocca\\
Via R. Cozzi 53, I-20126 Milano, Italy\\
paolo.lorenzoni@unimib.it\\[2ex]
}
\maketitle
\vspace{.2 cm}
\abstract{In this paper we consider a  class of semihamiltonian systems
 characterized by the existence of a special conservation law.
 The  density and the current of this conservation law 
 satisfy a second order system of PDEs which has a natural interpretation
 in the theory of flat bidifferential ideals. The class of systems we consider
 contains important well-known examples of semihamiltonian systems. Other 
 examples, like genus 1 
 Whitham modulation equations for KdV, are related to this 
 class by  a reciprocal trasformation. 
 }

\section{Introduction}
Bidifferential ideals play an important role 
 in the theory of finite-dimensional integrable systems, in particular in the
 bihamiltonian theory of separation of variables \cite{Ma,FP}.

Some recent results \cite{DM,plfm}
 suggest that they have also some applications 
 in the theory of infinite-dimensional integrable systems, in particular  
 in the case of integrable quasilinear PDEs. 

In this paper, following \cite{plfm}, we want to deepen the study
 of these applications in the case
 of diagonal integrable systems of quasilinear PDEs, the so called
 semihamiltonian systems. 

\begin{de} \cite{ts}
A diagonal system of PDEs of hydrodynamic type 
\begin{equation}
\label{hts}
u^i_t=v^i(u)u^i_x\hspace{1 cm}i=1,...,n
\end{equation}
is called semihamiltonian if the coefficients $v^i(u)$ satisfy 
the system of equations
\begin{equation}
\label{sh}
\partial_j\left(\frac{\partial_k v^i}{v^i-v^k}\right)=
\partial_k\left(\frac{\partial_j v^i}{v^i-v^j}\right)\hspace{1 
cm}\forall i\ne j\ne k\ne i,
\end{equation}
where $\d_i=\f{\partial}{\partial u^i}$. 
\end{de}

The equations (\ref{sh}) are the integrability conditions for the system
\begin{equation}
\label{sym} 
\f{\d_j w^i}{w^i-w^j}=\f{\d_j v^i}{v^i-v^j},
\end{equation}
which provides  the characteristic velocities $w^i$
 of the symmetries of (\ref{hts}), and for the system
\begin{equation}
\label{cl}
(v^i-v^j)\d_i\d_j H=\d_i v^j\d_j H-\d_j v^i\d_i H,
\end{equation}
which provides   the densities $H$  of conservation laws of (\ref{hts}). 

The knowledge of the symmetries of the system (\ref{hts}) allows one to
 find its general solution.  
Indeed, according to a general scheme of integration
 of semihamiltonian systems  proposed by Tsarev,
 the generalized hodograph method \cite{ts},
any solution of a semihamiltonian system is implicitly defined  by
 a system of algebraic equations
\begin{eqnarray}
\label{hm}
 w^i(u)=x+v^i(u) t\hspace{1 cm}i=1,...,n
\end{eqnarray}
 where the functions $w^i(u)$ are the solutions of the system (\ref{sym}).

A classical result in the theory of first order quasi-linear
 PDEs \cite{Lax} states that, if 
system (\ref{hts}) possess a conservation law 
\begin{equation*}
\d_t H+\d_x K=0
\end{equation*}
then the characteristic velocities $v^i$ can be written in the form
\begin{equation}
\label{sform}
v^i=-\f{\d_i K}{\d_i H}\hspace{1 cm}i=1,...,n.
\end{equation}
This result has some interesting consequences
 in the case of a semihamiltonian systems.
 Due to  integrability conditions (\ref{sh}) the space
 of solutions $w^i$ of the system
 (\ref{sym}) is parametrized by $n$ arbitrary functions of one variable. 

Since the system (\ref{cl}) is invariant with respect
 to the substitution $v^i\to w^i$,
 for any solution $(w^1,\dots,w^n)$ of the system (\ref{sym})
 there exists a function $K'$ such that
\begin{equation}
\label{sform2}
w^i=-\f{\d_i K'}{\d_i H}\hspace{1 cm}i=1,...,n.
\end{equation}
In other words the characteristic velocities of the symmetries 
  can be obtained applying the linear operator
\begin{equation}
\label{cvsh}
v^i_H(\cdot):=-\f{1}{\d_i H}\d_i(\cdot)\hspace{1 cm}i=1,...,n,
\end{equation} 
to a suitable current $K'$. 

Note that, in terms of the density $H$ and of the currents $K$ and $K'$,
 the system of algebraic equations (\ref{hm}) reads
\begin{equation*}
d(K'+xH-tK)=0.
\end{equation*}
Substituting  (\ref{sform2}) in (\ref{sym})
 and taking into 
account (\ref{cl}) we obtain the equations for the currents:
\begin{equation}
\label{current}
\d_i\d_j K'=\f{\d_j H}{\d_i H}\f{\d_i v^j}{v^i-v^j}\d_i K'
-\f{\d_i H}{\d_j H}\f{\d_j v^i}{v^i-v^j}\d_j K'.
\end{equation}

In general, the problem of finding the solutions
 of the system (\ref{current}) 
could be very difficult. The aim of the present paper
 is to study a special class of
 semihamiltonian systems characterized by the existence of 
 a density of conservation law $H$ such that  the equations (\ref{current})
 for the associated currents reduce to the form
\begin{equation}
\label{current2}
(f^i-f^j)\d_i\d_j K'=\d_i g^i\d_j K'-\d_j g^j\d_i K',
\end{equation}
where $f^i=f^i(u^i)$ and $g^i=g^i(u^i)$.

Surprinsingly, 
 also the density $H$ is a solution of the system (\ref{current2}).
 Therefore the solutions of (\ref{current2}) play a double role:

- fixed $H$, they are in one-to-one correspondence with the symmetries 
 of a semihamiltonian system (see formula (\ref{sform2})). In other words 
 they define a semihamiltonian hierarchy.

- they label these hierarchies: different choices of $H$ correspond to  
 different hierarchies.

The theory of flat bidifferential ideals arises naturally in this framework. 
 First of all because  any solution of the  
 system (\ref{current2}) defines a flat bidifferential ideal.
 Second because it
 provides a recursive procedure to compute the solutions of (\ref{current2}).

The paper is organized as follows:  in section 2 we recall 
some useful results about the theory of bidifferential ideals.
 In section 3 we apply these results
 to the theory of semihamiltonian systems.
 Section 4 is devoted to a discussion of the Hamiltonian formalism.
 In particular we find  a class of metrics
 satisfying a system of Egoroff-Darboux type. Remarkably, in general,
 these metrics are not related 
 to any Frobenius manifold, since their rotation coefficients are not
 symmetric.
Finally, in section 5, we put reciprocal transformations into the game.

\vspace{.5 cm}

{\bf Aknowledgements}. I wish to thank B. Dubrovin, G. Falqui,
 F. Magri and M. Pedroni for useful 
discussions. I am also grateful to M. Pavlov for interesting comments. 
This work has been partially supported 
by the European Community through the FP6 Marie Curie RTN {\em ENIGMA}
(Contract number MRTN-CT-2004-5652) and by (ESF) Scientific Programme
 {\em MISGAM}.

\section{Bidifferential ideals}

A tensor field $L:T{M}\rightarrow T{M}$,
 of type $(1,1)$ on a manifold $M$, 
of dimension $n$, is torsionless if the following identity
\begin{eqnarray*}
[LX,LY]-L[LX,Y]-L[X,LY]+L^2[X,Y]=0
\end{eqnarray*}
is verified for any pair of vector fields $X$ and $Y$ on $M$.  
According to the theory of graded derivations of Fr\"{o}licher-Nijenhuis 
 \cite{FN}, a torsionless tensor field $L$ of type $(1,1)$ defines a 
 differential operator $d_L$, of degree 1 and type 
$d$, on the Grassmann algebra of differential forms on $M$, verifying the 
fundamental conditions
\begin{eqnarray*}
d\cdot d_L+d_L\cdot d=0\hspace{2cm}d_L^2=0.
\end{eqnarray*}
On functions and 1-forms this derivation is defined by the following 
equations 
\begin{eqnarray*}
&&d_L f(X)=df(LX)\\
&&d_L\alpha(X,Y)=Lie_{LX}(\alpha(Y))-Lie_{LY}(\alpha(X))-\alpha([X,Y]_L),
\end{eqnarray*}
where
\begin{eqnarray*} 
[X,Y]_L=[LX,Y]+[X,LY]-L[X,Y].
\end{eqnarray*}
For instance, if $L=diag(f^1(u^1),\dots,f^n(u^n))$, the action of $d_L$ 
 on functions is given by the following formula:
\begin{equation*}
d_L g:=\sum_{i=1}^n f^i\f{\d g}{\d u^i}du^i
\end{equation*}
We can now define  the concept of bidifferential ideal of forms. 

\begin{de}
A bidifferential ideal $\mathfrak{I}$ is an ideal of differential forms 
on 
$M$ which is closed with respect to the action of both $d$ and $d_L$:
\begin{eqnarray*}
d(\mathfrak{I})\subset\mathfrak{I}\hspace{3 
cm}d_L(\mathfrak{I})\subset\mathfrak{I}
\end{eqnarray*}
\end{de}

For instance, if the ideal $\mathfrak{I}$ is generated by a single 
 1-form $\alpha$, the condition of closure with respect to the action
 of $d$ and $d_L$ reads
\begin{eqnarray*}
d\alpha=\lambda\wedge\alpha,\hspace{2 cm}d_L\alpha=\mu\wedge\alpha,
\end{eqnarray*}
where $\lambda$ and $\mu$ are suitable 1-forms.

In this paper we need a special subclass of bidifferential ideals, called
 flat bidifferential ideals.

\begin{de}
A flat bidifferential ideal $\mathfrak{I}$, of rank 1, on a manifold $M$ 
endowed with a torsionless tensor field $L:TM\rightarrow TM$, is the ideal 
of forms generated by the differential $dh$ of a function $h:M\rightarrow
 \mathbb{R}$ 
 obeying the condition
\begin{eqnarray}
\label{ddlh}
&&dd_L h=dh\wedge da
\end{eqnarray}
with respect to a function $a$ which satisfies
 the cohomological condition 
\begin{equation}
\label{ddla}
dd_L a=0.
\end{equation}
\end{de}
\begin{rem}
In the language of Dimakis and  M\"uller-Hoissen, 
the pair $(d,d_L+da\wedge)$ defines a ''gauged bidifferential calculus''.
 Some applications of this calculus to the theory 
 of integrable systems are discussed in \cite{DM}. 
\end{rem}

From now on, if not stated otherwise,
 we assume that the eigenvalues of $L$ are pairwise distinct.
 In this case the general solution of the equation (\ref{ddla}) is 
 given by the sum of $n$ arbitrary functions of one variable:
\begin{equation*}
a=\sum_{i=1}^n g^i(u^i)
\end{equation*}
and the cohomological equation (\ref{ddlh}) reads
\begin{equation}
\label{DDLH}
(f^i-f^j)\d_i\d_j h=\d_j h\d_i g^i-\d_i h\d_j g^j
\end{equation}
where $f^i(u^i)$ and $g^i=g^i(u^i)$ are arbitrary functions of one variable.

\section{Flat bidifferential ideals and semihamiltonian systems}

In this section
 we show that the linear operator (\ref{cvsh}) establishes a one-to-one
 correspondence
 between the space of solutions of the cohomological equation (\ref{ddlh})
 (which is 
parametrized by $n$ arbitrary functions of one variable) and the space of
 symmetries
 of a semihamiltonian system. Indeed, it is easy to prove the
 following proposition.

\begin{pro}
Let $H(u)$ be a  solution of the cohomological
 equation (\ref{ddlh}),  then:

\n
1)  the systems
\begin{eqnarray}
\label{fl1}
&&u^i_t=v^i_H(K_1)u^i_x,\hspace{1 cm}i=1,\dots,n,\\
\label{fl2}
&&u^i_{\tau}=v^i_H(K_2)u^i_x,\hspace{1 cm}i=1,\dots,n 
\end{eqnarray}
 commute for any pair ($K_1$,$K_2$) of solutions of (\ref{ddlh}).

\n
2) the  system of quasilinear PDEs
\begin{equation}
\label{exF}
u^i_t=[v^i_H(K)]u^i_x=\left[-\f{\d_i K}{\d_i H}\right]u^i_x,\hspace{1 cm}i=1,\dots,n
\end{equation}
is semihamltonian for any solution $K$ of the equation (\ref{ddlh}).
\end{pro}

\n
{\bf Proof}.

\n
1) The commutativity condition for the systems (\ref{fl1}) and (\ref{fl2})
 reads:
\begin{equation*}
\f{\d_j v^i_H(K_1)}{v^i_H(K_1)-v^j_H(K_1)}=
\f{\d_j v^i_H(K_2)}{v^i_H(K_2)-v^j_H(K_2)}.
\end{equation*}
By straightforward computation we get:
\begin{eqnarray}
\label{q}
\f{\d_j v^i_H(K)}{v^i_H(K)-v^j_H(K)}=
\f{\d_j H}{\d _i H}\f{\d_i\d_j H\d_i K-\d_i H\d_i\d_j K}
{\d_j K\d_i H-\d_i K\d_j H}=
\f{\d_i a}{f^j-f^i}\f{\d_j H}{\d_i H},
\end{eqnarray}
which does not depend on $K$.

\n
2)  By definition of semihamiltonian system we have to check that the 
 characteristic velocities $v^i_H(K)$ satisfy the system  (\ref{sh}).   
For $i\ne k\ne j\ne i$, we obtain the identity:
\begin{eqnarray*}
&&\d_k\left(\f{\d_j v^i_H(K)}{v^i_H(K)-v^j_H(K)}\right)=
\d_k\left(\f{\d_i a}{f^j-f^i}\f{\d_j H}{\d_i H}\right)\\
&&-\f{\d_i a}{(\d_i H)^2}
\left[\f{\d_j a\d_i H\d_k H}{
(f^i-f^j)(f^j-f^k)}
+\f{\d_k a\d_i H\d_j H}{(f^k-f^i)(f^j-f^k)}
+\f{\d_i a\d_j H\d_k H}
{(f^i-f^j)(f^k-f^i)}\right],
\end{eqnarray*}
which is clearly symmetric w.r.t. the indices $j$ and $k$.

\begin{flushright}
$\Box$
\end{flushright}

We have constructed a family of semihamiltonian systems  depending 
 on functional parameters: the eigenvalues $f^i(u^i)$ of $L$, the functions
 $a$ and $H$. If $\d_i f_i\ne 0,\,(i=1,\dots,n)$, without loss of generality, we can assume 
 $f^i(u^i)=u^i$; the two cases being simply related by the change of coordinates $u^i\to f^i(u^i)$.
 
Clearly, in order to make effective the construction one has to solve 
the cohomological equation (\ref{ddlh}).
 Even if its general
 solution, depending on $n$ arbitrary functions of one variable, is known explicitly
 only in some special
 cases (see section $6$ of \cite{Pv} and references therein), the  
  double differential complex defined by the pair $(d,d_L)$ allows one to construct iteratively
 a countable set of solutions.
  
\begin{lem}
Let $K_0$ be a solution of $(\ref{ddlh})$. Then,  
the functions $K_l$ defined recursively by 
\begin{equation}
\label{rec}
d K_{l+1}=d_L K_{l}-K_l da,
\end{equation}
satisfy the equation (\ref{ddlh}).
\end{lem}
The proof is based on standard arguments in the theory of
 bidifferential ideals \cite{Ma,DM}. We report it for convenience of
 the reader.

Let us start with the first step of the recursive procedure
\begin{equation}
\label{FS}
d K_{1}=d_L K_{0}-K_0 da,
\end{equation}
First of all, let us verify  that the 1-form appearing in 
 the right hand side of (\ref{FS}) is closed.
 Indeed, since $K_0$ is a solution of (\ref{ddlh}), applying 
to the right hand side of (\ref{FS}) the differential $d$ we obtain  
\begin{eqnarray*}
d\left(d_L K_0-K_0 da\right)=dK_0\wedge da-dK_0\wedge da=0.
\end{eqnarray*}
So the function $K_1$ is (locally) well defined. Moreover
\begin{eqnarray*}
dd_L K_1=d_L K_0\wedge da=dK_1\wedge da.
\end{eqnarray*}
We prove now the theorem by induction. Suppose that 
\begin{eqnarray*}
&&d K_{l}=d_L K_{l-1}-K_{l-1} da\\
&&dd_L K_l=dK_l\wedge da.   
\end{eqnarray*}
Then the 1-form in the right hand side of (\ref{rec}) is closed:
\begin{equation*}
d\left(d_L K_l-K_l da\right)=dK_l\wedge da-dK_l\wedge da=0	
\end{equation*}
and satisfies the equation (\ref{ddlh}):
\begin{eqnarray*}
dd_L K_{l+1}=d_L K_l\wedge da=dK_{l+1}\wedge da.
\end{eqnarray*}
\begin{flushright}
$\Box$
\end{flushright}

Let us illustrate how to apply the previous procedure in the case $H=a$, 
 $K_0=-a$. Using the recursive relations (\ref{rec}) we get
\begin{eqnarray*}
u^i_{t_0} &=& -\f{\d_i K_0}{\d_i a}u^i_x=u^i_x\\
u^i_{t_1} &=& -\f{\d_i K_1}{\d_i a}u^i_x=[f^i-a]u^i_x=[f^i+K_0]u^i_x\\
u^i_{t_2} &=& -\f{\d_i K_2}{\d_i a}u^i_x=[(f^i)^2+K_0 f^i+K_1]u^i_x\\
          &\vdots& \\  
u^i_{t_n} &=& -\f{\d_i K_n}{\d_i a}u^i_x=[(f^i)^n+K_0(f^i)^{n-1}+K_1(f^i)^{n-2}+\dots+K_{n-1}]u^i_x
\end{eqnarray*}

Following \cite{plfm} we can write the above hierarchy in the coordinate-free
 form:
\begin{eqnarray*}
u_{t_0} &=& u_x\\
u_{t_1} &=& [L+K_0 E]^i u_x\\
u_{t_2} &=& [L^2+K_0 L+K_1 E] u_x\\
          &\vdots& \\  
u_{t_n} &=& [L^n+K_0 L^{n-1}+K_1 L^{n-2}+\dots+K_{n-1}E] u_x
\end{eqnarray*}
where $u$ is the column vector $(u^1,\dots,u^n)^t$,
 $E$ is the identity matrix and $L$ is a torsionless tensor field
 of type $(1,1)$.
 The above vector fields commute also in the non 
 diagonalizable case \cite{plfm}. 

\begin{ex}
$H=a$, $L=diag(u^1,\dots,u^n)$, $a=c\,Tr(L)$  
\begin{eqnarray*}
&&K_0=-a=-c\sum_{j}u^j\\
&&K_1=-\frac{1}{2}c\sum_j (u^j)^2+
\frac{1}{2}c^2\left(\sum_j u^j\right)^2\\
&&K_2=-\frac{c}{3}\sum_j (u^j)^3+
\frac{c^2}{2}\sum_j (u^j)^2\sum_j u^j
-\frac{c^3}{6}\left(\sum_j u^j\right)^3
\end{eqnarray*}
and so on.
\end{ex}

\begin{ex} (Non diagonalizable case). 
Let
\begin{eqnarray*}
L=
\begin{bmatrix}
u^3 & \f{u^2}{2} & 0      \cr
0      & u^3 & \f{u^2}{2} \cr
0      & 0   & u^3 
\end{bmatrix}
\end{eqnarray*}
The function $a=u^1 (u^2)^2$ satisfies the cohomological equation
 $dd_L a=0$. Therefore the first non trivial flow of the hierarchy 
starting from $K_0=-a$ is 
\begin{eqnarray*}
\begin{bmatrix}
u^1_t \cr
u^2_t \cr
u^3_t 
\end{bmatrix}
=
\begin{bmatrix}
u^3-a  & \f{u^2}{2} & 0      \cr
0      & u^3-a & \f{u^2}{2} \cr
0      & 0   & u^3-a
\end{bmatrix}
\begin{bmatrix}
u^1_x \cr
u^2_x \cr
u^3_x 
\end{bmatrix}
\end{eqnarray*}
The other non trivial flows can be obtained solving the recursive 
 relations  (\ref{rec}) for the functions $K_1,K_2,\dots$:
\begin{eqnarray*}
K_1 &=& -u^1(u^2)^2 u^3-\f{1}{8}(u^2)^4+\f{1}{2}(u^1)^2(u^2)^4\\
K_2 &=& -u^1(u^2)^2 (u^3)^2-\f{1}{4}(u^2)^4 u^3+(u^1)^2(u^2)^2 u^3+\\
&&-
\f{1}{6}(u_1)^3(u_2)^6+\f{1}{8}u_1 (u_2)^6
\end{eqnarray*}
and so on.
\end{ex}

\begin{rem}
Semihamiltonian systems of the form
\begin{equation}
\label{pavsys}
u^i_{t} = (f^i-a)u^i_x,
\end{equation}
have been obtained in \cite{Pv} as finite component reduction of an infinite 
hydrodynamic chain.  The connection between bidifferential
 ideals and such systems 
has been investigated in \cite{plfm}.
 The starting point of that paper was the observation that the conditions 
(\ref{cl}) and (\ref{sh}) for systems (\ref{pavsys})
 coincide with the cohomological equations
\begin{eqnarray*}
&&dd_L H=da\wedge dH\\
&&dd_L a=0.
\end{eqnarray*}
 \end{rem}

\section{Some remarks about the Hamiltonian structure}

The Hamiltonian formalism for systems of
 hydrodynamic type was introduced by Dubrovin and Novikov 
in \cite{DN1,DN2}. They considered first order differential operators of the form
\begin{equation}
\label{PB}
P^{ij}=g^{ij}(u)\d_x-g^{is}\Gamma^{j}_{sk}(u)u^k_x 
\end{equation}
and the associated Poisson brackets
\begin{equation}
\label{PBHT}
\{F,G\}:=\int\f{\delta F}{\delta u^i}P^{ij}\f{\delta G}{\delta u^j}dx
\end{equation}
where $F=\int g(u)dx$ and $G=\int g(u)dx$ are local functionals.
\begin{teo}\cite{DN1}
If $\det{g^{ij}}\ne 0$, then the formula (\ref{PBHT}) 
 defines a Poisson bracket if and only if 
 the tensor $g^{ij}$ defines a flat pseudo-riemannian metric and the coefficients 
 $\Gamma^j_{sk}$ are the Christoffel symbols of the associated Levi-Civita connection.
\end{teo}
Non local extensions of the bracket (\ref{PBHT}), related to metrics of
 constant curvature, were considered by Ferapontov and Mokhov in \cite{MF}.
 Further generalizations 
 were considered by Ferapontov in \cite{Fera1}.

Let us focus our attention on semihamiltonan systems (\ref{hts})
\begin{equation*}
u^i_t=v^i(u)u^i_x,\hspace{1 cm}i=1,\dots n.
\end{equation*}
 In \cite{Fera1} Ferapontov conjectured that 
 any semihamiltonian system is always Hamiltonian with respect to suitable,
 possibly non local, Hamiltonian operators.
 Moreover he proposed the following construction to define
 such  Hamiltonian operators:

1. Find the general solution  of the system
\begin{equation}
\label{meq}
\d_j\ln{\sqrt{g_{ii}}}=\f{\d_j v^i}{v^j-v^i}.
\end{equation}
To this purpose is sufficient to find one solution $g_{ii}$  of (\ref{meq}). Indeed, 
the general solution is  $\f{g_{ii}}{\varphi^i(u^i)}$, where $\varphi^i$ are arbitrary functions
 of one argument. The flat solutions of (\ref{meq}) provide the local Hamitonian
 structures of the system (\ref{hts}).

2. Write the non vanishing components of the curvature tensor in terms
 of solutions $w_{\alpha}^i$ of the linear system (\ref{sym}):
\begin{equation}
\label{exp}
R^{ij}_{ij}=\sum_{\alpha}\epsilon_\alpha w^i_{\alpha}w^j_{\alpha}\hspace{1 cm}\epsilon_{\alpha}=\pm 1.
\end{equation}
(Ferapontov conjectured that it is always possible to find the expansion
 (\ref{exp})).

Then the system (\ref{hts}) is automatically Hamiltonian with respect to the
 Hamiltonian operator
\begin{equation}
\label{NLPB}
P^{ij}=g^{ii}\delta^{ij}\d_x-g^{ii}\Gamma^{j}_{ik}(u)u^k_x
+\sum_{\alpha}\epsilon_\alpha w^i_{\alpha}u^i_x\d_x^{-1}w^j_{\alpha}u^j_x.
\end{equation} 

\begin{ex}\cite{Fera1,tsarev,Pv2}
Let us consider the system of chromatography equations in Riemann invariants
\begin{equation}
\label{cheq}
u^i_t=\left[u^i\prod u^k\right]^{-1}u^i_x\hspace{1 cm}i=1,\dots n.
\end{equation}
The general solution of (\ref{meq}), in this case, is
\begin{equation*}
g_{ii}=\f{\prod_{k\ne i}(u^k-u^i)^2}{\varphi^i(u^i)}\hspace{1 cm}i=1,\dots n,
\end{equation*}
where $\varphi^i(u^i)$ are $n$ arbitrary functions of one variables.
 For $n\ge 3$ all these metrics 
 are not flat \cite{Pv2}. They generate nonlocal Hamiltonian operators
 of the form \cite{Fera1}
 \begin{equation}
\label{PBch}
P^{ij}=g^{ii}\delta^{ij}\d_x-g^{is}\Gamma^{j}_{sk}(u)u^k_x
-\sum_{\alpha=1}^n w^i_{\alpha}u^i_x\d_x^{-1}w^j_{\alpha}u^j_x,
\end{equation} 
where
\begin{equation}
w^i_1=\d_i\left(\f{\sqrt{\varphi^1}}{\prod_{l\ne 1}(u^l-u^1)^2}\right),\dots
w^i_n=\d_i\left(\f{\sqrt{\varphi^n}}{\prod_{l\ne n}(u^l-u^n)^2}\right).
\end{equation}
Note that system (\ref{cheq}) can be written in the form 
\begin{equation}
\label{sform3}
u^i_t=-\f{\d_i K}{\d_ i a}u^i_x,\hspace{1 cm}i=1,\dots n,
\end{equation}
where 
\begin{equation*}
K=-\f{1}{\prod_{k=1}^n u^k}
\end{equation*} 
is a solution of the cohomological equation (\ref{ddlh})
 with $L=diag(u^1,\dots u^n)$, $a=-Tr(L)$.
\end{ex}

\begin{ex}\cite{FPv}
Let us consider the semihamiltonian system
\begin{equation}
\label{sckdv}
u^i_t=\left[\sum_{i=1}^n u^i+2u^i\right]u^i_x,\hspace{1 cm}i=1,\dots n,
\end{equation}
The general solution of (\ref{meq}) is
\begin{equation*}
g_{ii}=\f{\prod_{k\ne i}(u^k-u^i)}{\varphi^i(u^i)},\hspace{1 cm}i=1,\dots n,
\end{equation*}
where $\varphi^i(u^i)$ are $n$ arbitrary functions of one variables.
 The choice $\varphi^i(u^i)=
(u^i)^{\alpha}$ $(\alpha=0,\dots,n$) provides $n+1$ flat metrics.
 For generic $\varphi^i(u^i)$ 
the metric $g_{ii}$ is not flat and generates nonlocal Hamiltonian operator with infinite 
 nonlocal tail.
 Note that system (\ref{sckdv}) can be written in the form (\ref{sform3}) 
 where 
\begin{equation*}
K=\f{1}{4}\sum_j (u^j)^2+\f{1}{8}\left(\sum_j u^j\right)^2,
\end{equation*} 
is a solution of the cohomological equation (\ref{ddlh}) with $L=diag(u^1,\dots u^n), a=-\f{1}{2}Tr(L)$.
\end{ex}   

Let us consider  semihamiltonian systems of the form (\ref{exF}).  
Taking into account the equation (\ref{ddlh}),  the system (\ref{meq})
 reduces to: 
\begin{equation}
\label{eps2}
\f{1}{2}\d_i\ln{g_{jj}}=-\d_j\ln{\d_i H}-\f{\d_j a}{f^i-f^j}.
\end{equation}
From now on, in this section, we assume $a=c\,Tr(L)=c\sum_{j=1}^n f^j$. In this case the general solution
 of (\ref{eps2}) is
\begin{equation} 
\label{sol}
g_{ii}=\f{(\d_i H)^2}{\varphi^i(u^i)[\prod_{l\ne i}(f^i-f^l)]^{2c}},
\hspace{1 cm}i=1,\dots n,
\end{equation}
where $\varphi^i(u^i)$ are $n$ arbitrary functions of one variable.

The rotation coefficients of the metrics (\ref{sol}) depend on the constant 
 $c$, the eigenvalues $(f^1,\dots,f^n)$ and on the choice of the arbitrary
 functions $(\varphi_1,\dots,\varphi_n)$ but not on the function $H$. More 
precisely we have the following proposition.

\begin{pro}
Let $H$ be a solution of the system:
\begin{equation}
\label{ddlhc}
(f^i-f^j)\d_i\d_j H=c\d_i f^i\d_j H-c\d_j f^j \d_i H
\end{equation}
Then, if $a=c Tr(L)$, the rotation coefficients  
$\beta_{ij}(u)=\f{\d_i\sqrt{g_{jj}(u)}}{\sqrt{g_{ii}(u)}}$ of the metrics (\ref{sol}),  does not depend on $H$. 
 More precisely they are given by the following expression
\begin{equation}
\label{rotcoef}
\beta_{ij}=
\left[\f{\prod_{l\ne i}(f^i-f^l)}{\prod_{l\ne j}(f^j-f^l)}\right]^c
\f{c\d_j f^j}{f^j-f^i}
\sqrt{\f{\varphi_i}{\varphi_j}}.
\end{equation}
\end{pro}

\n
{\bf Proof.}

\begin{eqnarray*}
&&\beta_{ij}(u)=\sqrt{\f{\varphi_i}{\varphi_j}}
\left[\f{\prod_{l\ne i}(f^i-f^l)}{\prod_{l\ne j}(f^j-f^l)}\right]^c
\left[\f{\d_i\d_j H}{\d_i H}+c\f{\d_j H}{\d_i H}\f{\d_j f_j}{f^j-f^i}\right].
\end{eqnarray*}
Taking into account (\ref{ddlhc}), we obtain formula (\ref{rotcoef}).

\begin{flushright}
$\Box$
\end{flushright}

The problem of finding the expansion (\ref{exp}) for the metrics (\ref{sol})
 is, in general, very difficult.

For $H=a$ this problem has been solved only for $c=\pm1,1/2$
 (see section 9 of \cite{Pv}).

The case $H\ne a$, which to our best knowledge, has not previously considered in the literature, 
 can be reduced to the case $H=a=c\,Tr(L)$. Indeed, we have the following proposition.

\begin{pro}
Let $H$ be a solution of the system 
\begin{equation*}
(f^i-f^j)\d_i\d_j H=c\d_i f^i\d_j H-c\d_j f^j\d_i H
\end{equation*}
 and 
\begin{eqnarray}
\label{Hhier}
u^i_{t_{\alpha}}=\tilde{w}^i_{\alpha}u^i_x=
-\f{\d_i K_{\alpha}}{\d_i H}u^i_x,
\end{eqnarray}
the corresponding semihamiltonian hierarchy constructed with the solutions 
 $K_{\alpha}$ of the system (\ref{ddlhc}).
 Suppose that the hierarchy 
\begin{equation}
\label{H=a}
u^i_{t_{\alpha}}=w^i_{\alpha}u^i_x
=-\f{\d_i K_{\alpha}}{c\d_i f^i}u^i_x
\end{equation}
 is Hamiltonian w.r.t. the  Hamiltonian operator 
\begin{equation}
\label{PBa}
P^{ij}=g^{ii}\delta^{ij}\d_x-g^{ii}\Gamma^{j}_{ik}(u)u^k_x
-\sum_{\alpha} w^i_{\alpha}u^i_x\d_x^{-1}w^j_{\alpha}u^j_x
\end{equation}
and the Hamiltonian densities $h_{\alpha}$. 
Then the hierarchy (\ref{Hhier}) is Hamiltonian w.r.t.
 the  Hamiltonian operator 
\begin{equation}
\label{PB'}
\tilde{P}^{ij}=\tilde{g}^{ii}\delta^{ij}\d_x-\tilde{g}^{ii}
\tilde{\Gamma}^{j}_{ik}(u)u^k_x
-\sum_{\alpha} \tilde{w}^{i}_{\alpha}u^i_x
\d_x^{-1}\tilde{w}^{j}_{\alpha}u^j_x,
\end{equation}
where 
$\tilde{g}^{ii}=\left[c\f{\d_i f^i}{\d_i H}\right]^2 g^{ii}$, the coefficients
 $\tilde{\Gamma}^{j}_{ik}$ are the Christoffel symbols of the associated 
 Levi-Civita connection and $\tilde{w}^{i}_{\alpha}
=\f{c\d_i f^i}{\d_i H}w^i_{\alpha}$. 
Moreover the Hamiltonian densities $\tilde{h}_{\alpha}$
 of the systems (\ref{Hhier}) can be obtained 
from the Hamiltonian densities $h_{\alpha}$ solving the compatible system
\begin{equation}
\label{hs}
\d_i \tilde{h}_{\alpha}=\f{\d_i H}{\d_ i a}\d_i h_{\alpha}.
\end{equation}
\end{pro}

\n
{\bf Proof:} the non vanishing components of the curvature tensor 
\begin{equation}
\label{rij}
R^{ij}_{ij}=
g^{ii}\left(\d_j\Gamma^j_{ii}-\d_i\Gamma^j_{ij}-\Gamma^j_{pi}\Gamma^p_{ij}+\Gamma^j_{pj}\Gamma^p_{ii}
\right)
\end{equation}
can be written in the form
\begin{equation}
\label{rij2}
R^{ij}_{ij}=\f{1}{\d_i H\d_j H}S^{ij}_{ij}
\end{equation}
where the quantities $S^{ij}_{ij}$ do not depend on $H$.
Indeed in terms of the rotation coefficients (that do not depend on $H$),
 formula (\ref{rij}) reads:
\begin{equation*}
R^{ij}_{ij}=-\f{1}{\sqrt{g_{ii}}}\f{1}{\sqrt{g_{jj}}}\left(\d_i\beta_{ij}
+\d_j\beta_{ji}+\sum_{k\ne i,j}\beta_{ki}\beta_{kj}\right).
\end{equation*}
Using this fact it is easy to obtain the expansion (\ref{exp}) 
for the non vanishing components of the curvature tensor
 of the metric  $g'_{ii}=c^2\left(\f{\d_i H}{\d_i f^i}\right)^2 g_{ii}$:
\begin{equation*}
R^{'ij}_{ij}=c^2\f{\d_i f^i\d_j f^j}{\d_i H\d_j H}R^{ij}_{ij}.
\end{equation*}

Observe that the coefficients $\tilde{w}^i_{\alpha}=c\f{\d_i f^i}{\d_i H}
w^i_{\alpha}=-\f{\d_i K_{\alpha}}{\d_i H}$ are characteristic velocities 
 of symmetries of (\ref{Hhier}). Therefore the bivector (\ref{PB'})
 satisfies all Ferapontov conditions. Indeed:

- the diagonal metric $\tilde{g}_{ii}$ is a solution of the system (\ref{meq}).

- the coefficients $\tilde{\Gamma}^i_{jk}$ are, by definition, the Christoffel
 symbols of the associated Levi-Civita connection.

- the non local tail of (\ref{PB'}) is constructed with  
  the characteristic velocities $\tilde{w}^i_{\alpha}$ appearing in 
 the expansion of the non vanishing components of the curvature tensor.

We have to show now that the function $\tilde{h}_{\alpha}$ are hamiltonian 
 densities. First of all we observe that they are well-defined. Indeed
 the compatibility of the system (\ref{hs}) reads
\begin{eqnarray}
\label{hal}
\d_i\d_j h_{\alpha}-c\f{\d_i h_{\alpha}\d_j f^j}{f^i-f^j}
+c\f{\d_j h_{\alpha}\d_i f^i}{f^i-f^j}=0,
\end{eqnarray}
which is nothing but the system (\ref{cl})
 for the densities of conservation law 
of the semihamiltonian hierarchy (\ref{H=a}). Moreover it is easy to check
 that if the functions $h_{\alpha}$ are solutions of the system (\ref{hal}),
 then  the functions $\tilde{h}_{\alpha}$ are solutions of the system
 (\ref{cl}) for the densities of conservation laws of the semihamiltonian
 hierarchy (\ref{Hhier}).

\begin{flushright}
$\Box$
\end{flushright}

We conclude this section mentioning an important property of the metrics (\ref{sol}) in the case
 $L=diag(u^1,\dots,u^n)$ and $\varphi_i=1$ $(i=1,\dots,n)$.

\begin{pro}
 If $f^i(u^i)=u^i$ and $\varphi_i=1$ $(i=1,\dots,n)$, 
 the rotation coefficients (\ref{rotcoef}) satisfy the system 
\begin{eqnarray}
\label{cv1}
&&\d_k\beta_{ij}=\beta_{ik}\beta_{kj}\hspace{1 cm}i\ne j\ne k\\
\label{cv2}
&&\sum_{k}\d_k\beta_{ij}=0\hspace{1 cm}i\ne j\\
\label{qo}
&&\sum_{k}u^k\d_k\beta_{ij}=-\beta_{ij}\hspace{1 cm}i\ne j
\end{eqnarray}
\end{pro}

\n
{\bf Proof:} the equations (\ref{cv1}) are automatically satisfied because they are
 equivalent to the conditions (\ref{sh}). 

%\begin{eqnarray*}
%\d_k\gamma_{ij}&=&\left[\f{\prod_{l\ne j}(f^j-f^l)}{\prod_{l\ne i}(f^i-f^l)}\right]^{c-1}
%\f{c^2\d_i f^i\d_j f^k}{f^i-f^j}\times\\
%&&\left[-\f{\prod_{l\ne j,k}(f^j-f^l)}{\prod_{l\ne i}(f^i-f^l)}
%+\f{\prod_{l\ne j}(f^j-f^l)\prod_{l\ne i,k}(f^i-f^l)}{[\prod_{l\ne i}(f^i-f^l)]^2}\right]=\\
%&&\left[\f{\prod_{l\ne j}(f^j-f^l)}{\prod_{l\ne i}(f^i-f^l)}\right]^{c}
%\f{c^2\d_i f^i\d_j f^k}{f^i-f^j}\left[\f{1}{f^k-f^j}+\f{1}{f^i-f^k}\right]=\\
%&&\left[\f{\prod_{l\ne j}(f^j-f^l)}{\prod_{l\ne i}(f^i-f^l)}\right]^{c}
%\f{c^2\d_i f^i\d_j f^k}{(f^k-f^j)(f^i-f^k)}=\gamma_{ik}\gamma_{kj}
%\end{eqnarray*}

Moreover, by straightforward computation we obtain
\begin{eqnarray*}
&&\sum_k \d_k\beta_{ij}=\d_j\beta_{ij}+\d_i\beta_{ij}+\sum_{k\ne i,j}\beta_{ik}\beta_{kj}=\\
&&\left[\f{\prod_{l\ne i}(u^i-u^l)}{\prod_{l\ne j}(u^j-u^l)}\right]^{c-1}
\f{c^2}{u^j-u^i}\times\\
&&\left\{
-\f{\prod_{l\ne i,j}(u^i-u^l)}{\prod_{l\ne j}(u^j-u^l)}-
\f{\prod_{l\ne i}(u^i-u^l)\sum_{k\ne j}\prod_{l\ne j,k}(u^j-u^l)}{\prod_{l\ne j}(u^j-u^l)^2}
+\right.\\
&&\left.
\f{\sum_{k\ne i}\prod_{l\ne i,k}(u^i-u^l)}{\prod_{l\ne j}(u^j-u^l)}+
\f{\prod_{l\ne i}(u^i-u^l)\prod_{l\ne i,j}(u^j-u^l)}{\prod_{l\ne j}(u^j-u^l)^2}
\right\}+\\
&&+\left[\f{\prod_{l\ne i}(u^i-u^l)}{\prod_{l\ne j}(u^j-u^l)}\right]^{c}
\f{c^2}{u^j-u^i}\sum_{i,j}\left[\f{1}{u^k-u^i}+\f{1}{u^j-u^k}\right]=\\
&&\left[
\f{\prod_{l\ne i}(u^i-u^l)}{\prod_{l\ne j}(u^j-u^l)}
\right]^{c}
\f{c^2}{u^j-u^i}\times\\
&&\left\{-\sum_{k\ne i,j}\f{1}{u^j-u^k}+\sum_{k\ne i,j}\f{1}{u^i-u^k}+\sum_{k\ne i,j}
\left(\f{1}{u^k-u^i}+\f{1}{u^j-u^k}\right)\right\}=\\
&&=0
\end{eqnarray*}
and
\begin{eqnarray*}
&&\sum_k u^k\d_k\beta_{ij}=\sum_{k\ne i,j}u^k\beta_{ik}\beta_{kj}+u^j\d_j\beta_{ij}+u^i\d_i\beta_{ij}=\\
&&\left[\f{\prod_{l\ne i}(u^i-u^l)}{\prod_{l\ne j}(u^j-u^l)}\right]^{c}
\left\{
\f{c^2 u^k}{u^j-u^i}\sum_{k\ne i,j}\left[
\f{1}{u^k-u^i}+\f{1}{u^j-u^k}\right]-\f{c}{u^j-u^i}+\right.\\
&&\left.\f{c^2}{u^j-u^i}\left[\f{u^j}{u^j-u^i}-\sum_{k\ne j}\f{u^j}{u^j-u^k}
+\sum_{k\ne i}\f{u^i}{u^i-u^k}+\f{u^i}{u^j-u^i}\right]\right\}=\\
&&-\beta_{ij}+\left[\f{\prod_{l\ne i}(u^i-u^l)}{\prod_{l\ne j}(u^j-u^l)}\right]^{c}
\f{c^2}{u^j-u^i}\times\\
&&\left\{
\sum_{k\ne i,j}\left[\f{u^k}{u^k-u^i}+\f{u^k}{u^j-u^k}\right]
-\sum_{k\ne i,j}\f{u^j}{u^j-u^k}+\sum_{k\ne i,j}\f{u^i}{u^i-u^k}
\right\}=\\
&&-\beta_{ij}+\left[\f{\prod_{l\ne i}(u^i-u^l)}{\prod_{l\ne j}(u^j-u^l)}\right]^{c}
\f{c^2}{u^j-u^i}\left\{\sum_{k\ne i,j}
\left[\f{u^k-u^j}{u^j-u^k}+\f{u^k-u^i}{u^k-u^i}\right]\right\}=\\
&&=-\beta_{ij}.
\end{eqnarray*}

\begin{flushright}
$\Box$
\end{flushright}

\begin{rem}  
In general, the rotation coefficients (\ref{rotcoef}) are not symmetric. 
In the case of symmetric rotation coefficients, 
   the equations (\ref{cv1}), (\ref{cv2}) and (\ref{qo})
 arise naturally in the framework of Frobenius manifolds \cite{Dub}.

As well-known the theory of
 Frobenius manifolds is related to the theory of isomonodromic
 deformations. 
 Indeed, 
the equations (\ref{cv1}), (\ref{cv2}) and (\ref{qo}) 
  are equivalent to the system:
\begin{equation}
\label{iso}
\d_k V(u)=[V(u),[E_k,\Gamma]],\hspace{1 cm}k=1,\dots,n,
\end{equation}
where
\begin{eqnarray*}
&&(E_k)_{ij}=\delta_{ik}\delta_{kj}\\
&&U:=diag(u^1,\dots,u^n)\\
&&\Gamma(u):=(\beta_{ij})\\
&&V(u):=(u^i-u^j)\beta_{ij},
\end{eqnarray*}
that governs the monodromy preserving deformations of the 
 operator 
\begin{equation*}
\f{d}{dz}-\left(U+\f{V}{z}\right).
\end{equation*}
The proof of this fact (see \cite{Dub}) does not rely on the hypothesis
 of symmetry of the rotation coefficients. 
\end{rem}

\section{A remark on Whitham equations}

A well-known example of semihamiltonian system is
 the system of quasilinear PDEs that describes the slow modulations 
of $g$-gap solutions of the KdV hierarchy: the Whitham equations \cite{wh}.
 In this case 
 the characteristic velocities can be written in terms of hyperelliptic
 integrals of genus $g$. In $g=1$ case these equations read:
\begin{eqnarray*}
&&u^1_t=\left[-\f{u^1+u^2+u^3}{3}+\f{2(u^2-u^1)K(s)}{3(K(s)-E(s)}\right]u^1_x\\
&&u^2_t=\left[-\f{u^1+u^2+u^3}{3}+\f{2(u^2-u^1)K(s)}{3(E(s)-(1-s^2)K(s)}\right]
u^2_x\\
&&u^3_t=\left[-\f{u^1+u^2+u^3}{3}-\f{2(u^3-u^1)(1-s^2)E(s)}{3E(s)}\right]u^3_x
\end{eqnarray*}
where $s=\f{u^2-u^1}{u^3-u^1}$, $K(s)$ and $E(s)$ are complete elliptic integrals
 of the first and second kind. 

For Whitham equations the hodograph method is effective
\cite{kr,KS,tian0,tian1,tian2,GKE1,GKE2,El,G}. Indeed
 it is possible to construct explicitly the symmetries appearing 
in the equations (\ref{hm}):
\begin{teo} \cite{KS,tian0,tian1,GKE1,GKE2}
There exist functions $q_1(u),q_2(u),q_3(u)$
 such that the characteristic velocities
 $w^i$ of the symmetries of the Whitham equations have the form
\begin{equation}
\label{mt}
w^i:=\left[1+q_i\d_i\right]K,\hspace{1 cm}i=1,...,3,
\end{equation}
where the function $K$ is a solution of the following system of Euler-Poisson-Darboux type:
\begin{equation}
\label{DS}
2(u^i-u^j)\d_i\d_j K=\d_i K-\d_j K\hspace{1 cm}i\ne j,\hspace{.2 cm}i,j=1,2,3.
\end{equation}
which can be explicitly solved.
\end{teo}
The functions $q_i(u)$ can be written in terms of the complete elliptic integral $K(s)$ 
 and $E(s)$. Moreover from the conservation of waves it follows that
 \cite{Ku,GKE1,GKE2,tian2}:
\begin{equation*}
q_i(u)=-\f{H}{\d_i H}\hspace{1 cm}i=1,2,3,
\end{equation*}
where  
\begin{equation}
H=\oint\f{d\xi}{\sqrt{(u^1-\xi)(u^2-\xi)(u^3-\xi)}}
\end{equation}
 is the wavelength 
(the integration is taken over the cycle around the gap $u^2\le\xi\le u^3$). 
Therefore, the Whitham equations can be written in the form:
\begin{equation*}
u^i_t=v^i_H(K)=\left[K-\f{H}{\d_i H}\d_i K\right]u^i_x,\hspace{1 cm}i=1,2,3.
\end{equation*}
Note  that  the wavelength $H$ satisfies 
 the Euler-Darboux-Poisson system (\ref{DS}),
 which is a particular case of the cohomological
 equation (\ref{ddlh}) corresponding to the choice
 $L=diag(u^1,u^2,u^3)$, $a=-\f{1}{2}Tr(L)$. This remark suggests 
 to consider systems of the form 
\begin{equation}
\label{MT}
u^i_t=v^i_H(K)u^i_x
=\left[K-\f{H}{\d_i H}\d_i K\right]u^i_x,\hspace{1 cm}i=1,...,n.
\end{equation}
 where the
 function $H$ and $K$ are solutions of the cohomological equations
 (\ref{ddlhc}). It is easy to prove that such systems are semihamiltonian 
 and that, fixed $H$, the systems
\begin{eqnarray*}
&&u^i_t=v^i_H(K_1)u^i_x,\hspace{1 cm}i=1,...,n\\
&&u^i_{\tau}=v^i_H(K_2)u^i_x,,\hspace{1 cm}i=1,...,n,
\end{eqnarray*}
 commute for any pair $(K_1,K_2)$ of (\ref{ddlhc}).

This fact can be proved by straightforward calculation or simply observing
 that systems of the form (\ref{MT}) can be obtained from the systems 
 studied in this paper by means of a reciprocal transformation.

\begin{pro}
Systems 
\begin{equation*}
u^i_{\tilde{t}}
=\left[K-\f{H}{\d_i H}\d_i K\right]u^i_{\tilde{x}},\hspace{1 cm}i=1,...,n.
\end{equation*}
are related to the systems 
\begin{equation*}
u^i_t=\left[-\f{\d_i K}{\d_i H}\right]u^i_x\hspace{1 cm}
\end{equation*}
by the reciprocal transformation
\begin{eqnarray*}
&&d\tilde{x}=Hdx-Kdt\\
&&d\tilde{t}=dt
\end{eqnarray*}
\end{pro}  

The proof is a trivial computation.

\section{Conclusions}
In this paper we studied some applications of the theory of
 flat bidifferential ideals to semihamiltonian systems
 of quasilinear PDEs.

The starting point of the present paper was the observation that for any
 semihamiltonian system there exists a linear differential operator that,
 acting on a suitable domain, 
  provides all the symmetries of the system.

We showed that for a special class of semihamiltonian systems 
 this operator 
 and its domain are completely characterized by the solutions
 of a cohomological equation.

Moreover the theory of
 flat bidifferential ideals
 naturally provides a recursive procedure to compute the solutions
 of this equation.

\end{document}